\documentclass[twocolumn,showpacs,preprintnumbers,amsmath,amssymb,superscriptaddress,prd]{revtex4}

\usepackage{graphicx}
\usepackage{dcolumn}
\usepackage{bm}
\usepackage{amsmath,amssymb}

\begin{document}

\preprint{APS/123-QED}

\title{Constraints on $R^n$ gravity from precession of orbits of S2-like
stars}

\author{D. Borka}

\email[Corresponding author:]{dusborka@vinca.rs} \affiliation{Atomic
Physics Laboratory (040), Vin\v{c}a Institute of Nuclear Sciences,
University of Belgrade, P.O. Box 522, 11001 Belgrade, Serbia}

\author{P. Jovanovi\'{c}}
\affiliation{Astronomical Observatory, Volgina 7, 11060 Belgrade,
Serbia}

\author{V. Borka Jovanovi\'{c}}
\affiliation{Atomic Physics Laboratory (040), Vin\v{c}a Institute of
Nuclear Sciences, University of Belgrade, P.O. Box 522, 11001
Belgrade, Serbia}

\author{A. F. Zakharov}
\affiliation{Institute of Theoretical and Experimental Physics, B.
Cheremushkinskaya 25, 117259 Moscow, Russia}

\affiliation{Bogoliubov Laboratory for Theoretical Physics, JINR,
141980 Dubna, Russia}

\date{5 june 2012}

\begin{abstract}
We study some possible observational signatures of $R^n$ gravity at
Galactic scales and how these signatures could be used for
constraining this type of $f(R)$ gravity. For that purpose, we
performed two-body simulations in $R^n$ gravity potential and
analyzed the obtained trajectories of S2-like stars around Galactic
center, as well as resulting parameter space of $R^n$ gravity
potential. Here, we discuss the constraints on the $R^n$ gravity
which can be obtained from the observations of orbits of S2-like
stars with the present and next generations of large telescopes. We
make comparison between the theoretical results and observations.
Our results show that the most probable value for the parameter
$r_c$ in $R^n$ gravity potential in the case of S2-like stars is
$\sim$100 AU, while the universal parameter $\beta$ is close to
0.01. Also, the $R^n$ gravity potential induces the precession of
S2-like stars orbit in opposite direction with respect to General
Relativity, therefore, such a behavior of orbits qualitatively is
similar to a behavior of Newtonian orbits with a bulk distribution
of matter (including a stellar cluster and dark matter
distributions).
\end{abstract}

\pacs{04.50.Kd, 04.80.Cc, 04.25.Nx, 04.50.-h}

\maketitle

\section{Introduction}

Power-law fourth-order theories of gravity have been proposed like
alternative approaches to Newtonian gravity \cite{capo06,capo07}. In
this paper we study possible application of $R^n$ gravity on
Galactic scales, for explaining observed precession of orbits of
S-stars, as well as weather these observations could be used for
constraining this type of $f(R)$ gravity \cite{soti10}.

S-stars are the bright stars which move around the massive black
hole in the center of our Galaxy \cite{gill09a,gill09b,ghez08}. For
one of them, called S2, there are some observational indications
that this orbit deviates from the Keplerian case due to relativistic
procession \cite{gill09a}. Besides, an extended dark mass which
probably exists in the Galactic center, could also contribute to
pericenter precessing of the S2, but in the opposite direction
\cite{Nucita_07,gill09a}. Progress in monitoring bright stars near
the Galactic Center have been made recently \cite{gill09a}. With the
Keck 10 m telescope, the several stars orbiting the black hole in
Galactic Center have been monitored, and in some cases almost entire
orbits, as, for example, that of the S2 star, have been observed,
allowing an unprecedented description of the Galactic Center region
\cite{gill09a}. The astrometric limit for S2 star orbit is today
around 10 mas and within that limit one can not say for sure that S2
star orbit really deviates from the Newtonian case. In the future,
it will be possible to measure the positions of some stars with
astrometric errors several times smaller than errors of current
observations and that is why we will consider here even smaller
astrometric limits.

Capozziello et al. \cite{capo07} investigated the possibility that
the observed flatness of the rotation curves of spiral galaxies is
not evidence for the existence of dark matter (DM) haloes, but
rather a signal of the breakdown of General Relativity (GR). They
found a very good agreement between the theoretical rotation curves
and the data using only stellar disc and interstellar gas when the
slope $n$ of the gravity Lagrangian is set to the value $n$ = 3.5
(giving $\beta$ = 0.817), obtained by fitting the Type Ia supernova
Hubble diagram with the assumed power-law $f(R)$ model and without
dark matter \cite{capo07}.

Frigerio Martins and Salucci \cite{fms07} have also investigated the
possibility of fitting the rotation curves of spiral galaxies with
the power-law fourth-order theory of gravity, without the need for
dark matter. They show that, in general, the power law {\bf $f(R)$}
version could fit the observations well, with reasonable values for
the mass model.

Recently, gravitational microlensing has been investigated in the
framework of the weak field limit of fourth order gravity theory
\cite{zakh06}. The solar system data (i.e. planetary periods) and
light bending due to microlensing can be used to put strong
constraints on the parameters of this class of gravity theories. In
paper \cite{zakh06} it was found that these parameters must be very
close to those corresponding to the Newtonian limit of the theory.
In  paper \cite{zakh07} the authors discuss the constraints that can
be obtained from the orbit analysis of stars (as S2 and S16) moving
inside the DM concentration. In particular, consideration of the S2
star apoastron shift may allow improving limits on the DM mass and
size.

Rubilar and Eckart \cite{rubi01} investigated the properties of
stellar orbits close to central mass and the corresponding
connection with current and (near) future observational
capabilities. They showed that the orbital precession can occur due
to relativistic effects, resulting in a prograde shift, and due to a
possible extended mass distribution, producing a retrograde shift.
Both, prograde relativistic and retrograde Newtonian periastron
shifts will result in rosette shaped orbits. Weinberg et al.
\cite{wein05} discuss physical experiments achievable via the
monitoring of stellar dynamics near the massive black hole at the
Galactic Center with a diffraction-limited, next-generation,
extremely large telescope (ELT). They demonstrate that the lowest
order relativistic effects, such as the prograde precession, will be
detectable if the astrometric precision become less then 0.5 mas.

In this paper we continue to investigate constraints on the
parameters of this class of gravity theories using S2-like star
orbits under uncertainty of 10 mas. In Section \S2 the type of used
gravitational potential is given. In Section \S3 we present the
S2-like stars orbits, gravity parameters and angles of orbital
precession, and also compared theoretical results with observations.
The main conclusions are pointed out in \S4.

\begin{figure*}[ht!]
\centering
\includegraphics[width=0.85\textwidth]{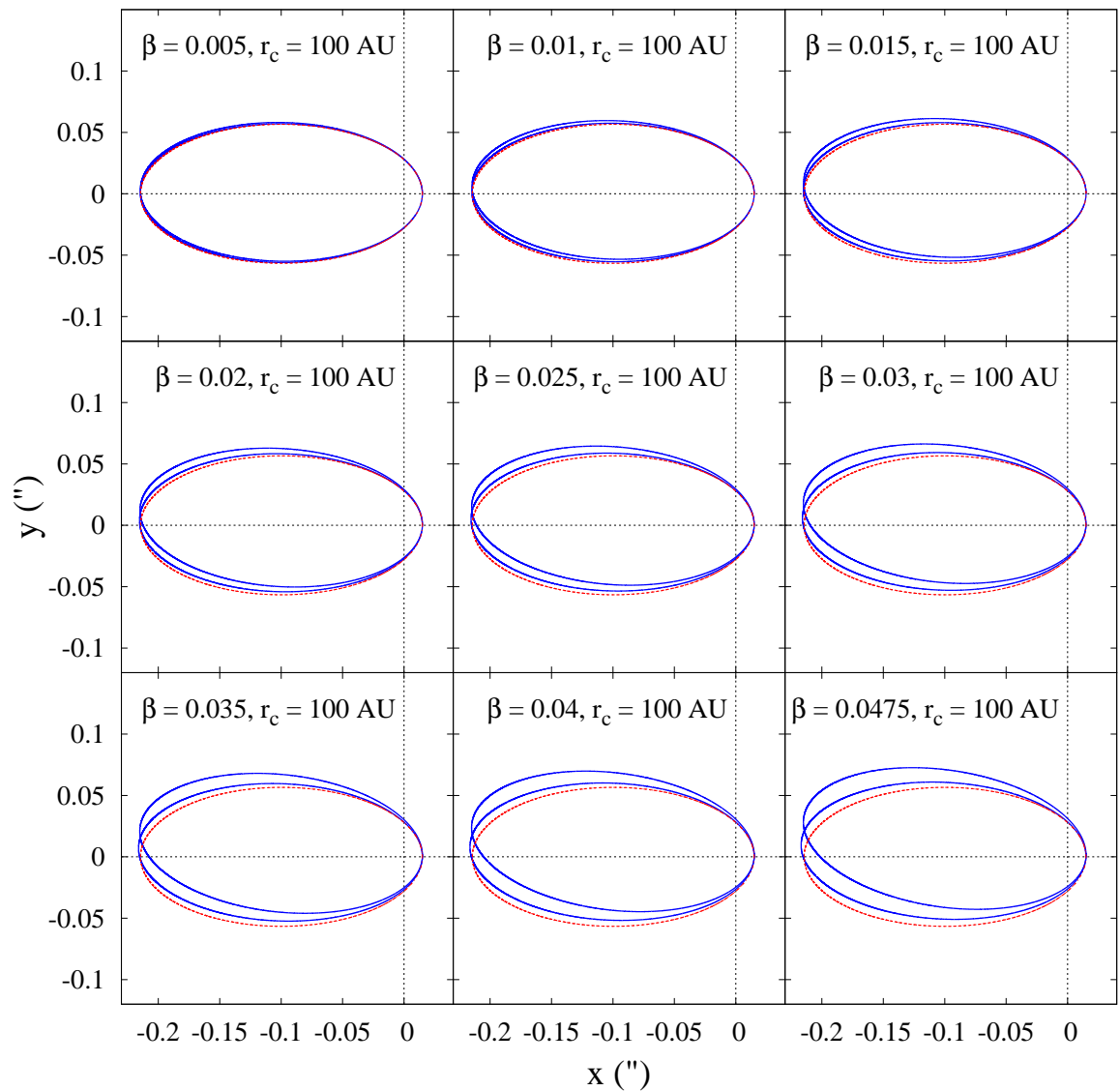}
\caption{The orbits of an S2-like star around massive black hole in
$R^n$ gravity (blue solid line) and in Newtonian gravity (red dashed
line) for $r_c$ = 100 AU and for these nine values of parameter
$\beta$: 0.005, 0.01, 0.015, 0.02, 0.025, 0.03, 0.035, 0.04, 0.0475.
The black hole with mass $M_{BH}=3.4\times 10^6\ M_\odot$ is assumed
to be located at coordinate origin, and mass of S2-like star is
taken to be $M_\star=1\ M_\odot$.}
\label{fig01}
\end{figure*}

\begin{figure*}[ht!]
\centering
\includegraphics[width=0.85\textwidth]{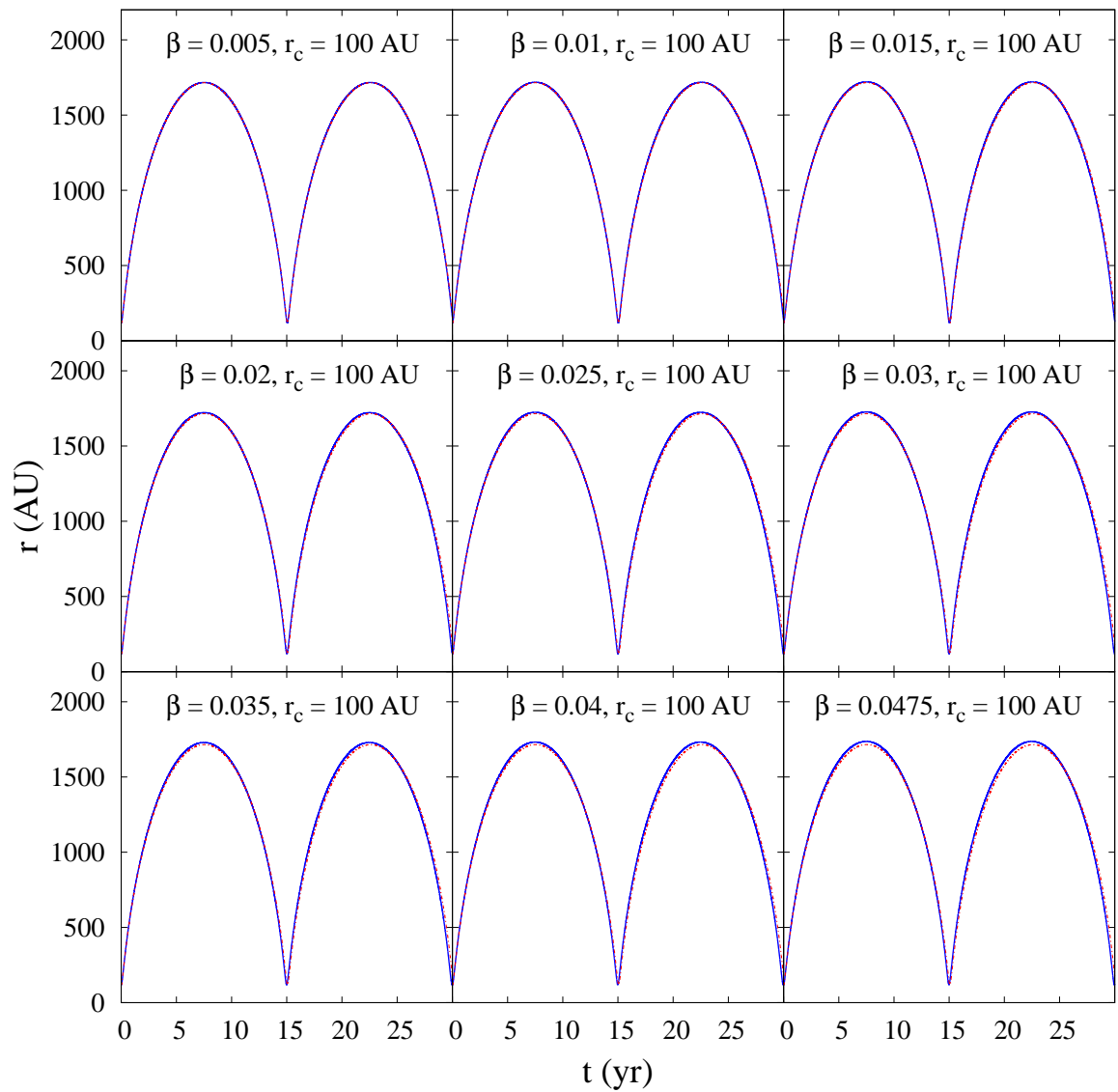}
\caption{The distances between the S2-like star and black hole as a
function of time for the same values of parameters $r_c$ and $\beta$
as in the Fig.
\ref{fig01}.}
\label{fig02}
\end{figure*}

\begin{figure*}[ht!]
\centering
\includegraphics[width=0.85\textwidth]{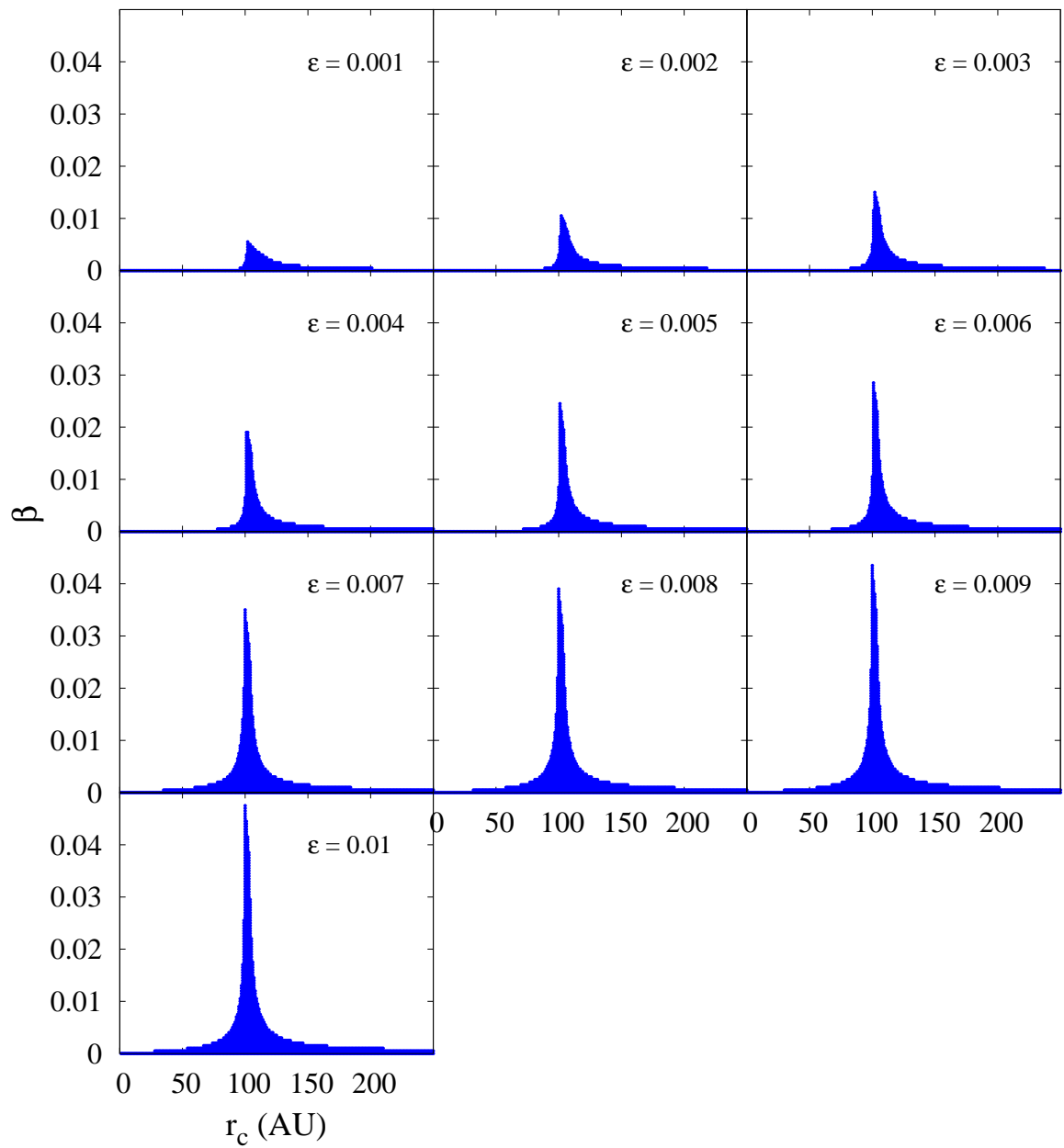}
\caption{The parameter space for $R^n$ gravity under the constraint
that, during one orbital period, S2-like star orbits in $R^n$
gravity differ less than $\varepsilon$ from the corresponding orbits
in Newtonian gravity, for the following 10 values of parameter
$\varepsilon$: 0.001, 0.002, 0.003, 0.004, 0.005, 0.006, 0.007,
0.008, 0.009 and 0$''$.01.}
\label{fig03}
\end{figure*}

\begin{figure*}[ht!]
\centering
\includegraphics[width=0.85\textwidth]{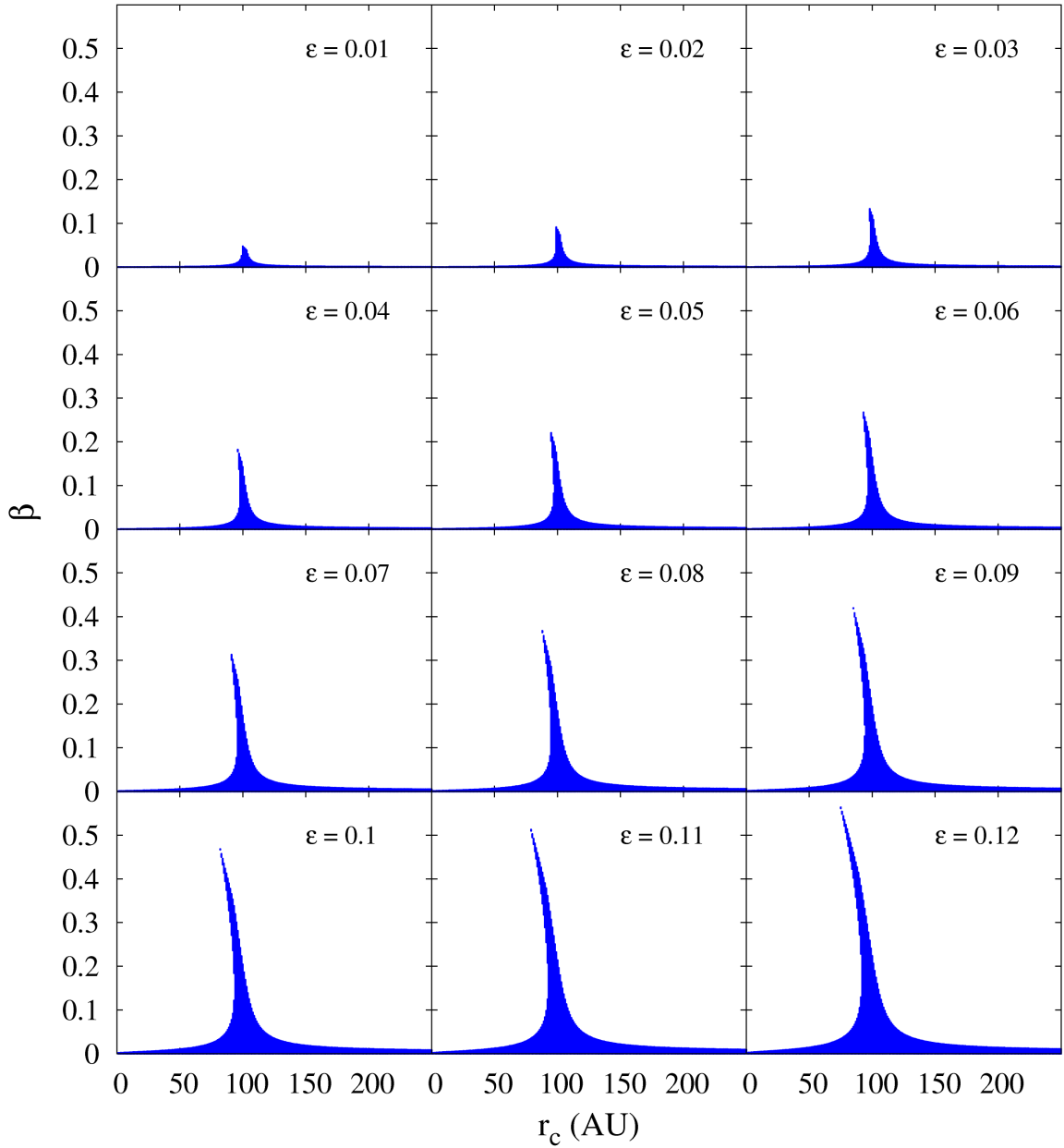}
\caption{The same as in Fig. \ref{fig03} but for the following 12
values of parameter $\varepsilon$: 0.01, 0.02, 0.03, 0.04, 0.05,
0.06, 0.07, 0.08, 0.09, 0.1, 0.11 and 0$''$.12.}
\label{fig04}
\end{figure*}

\begin{figure*}[ht!]
\centering
\includegraphics[width=0.40\textwidth]{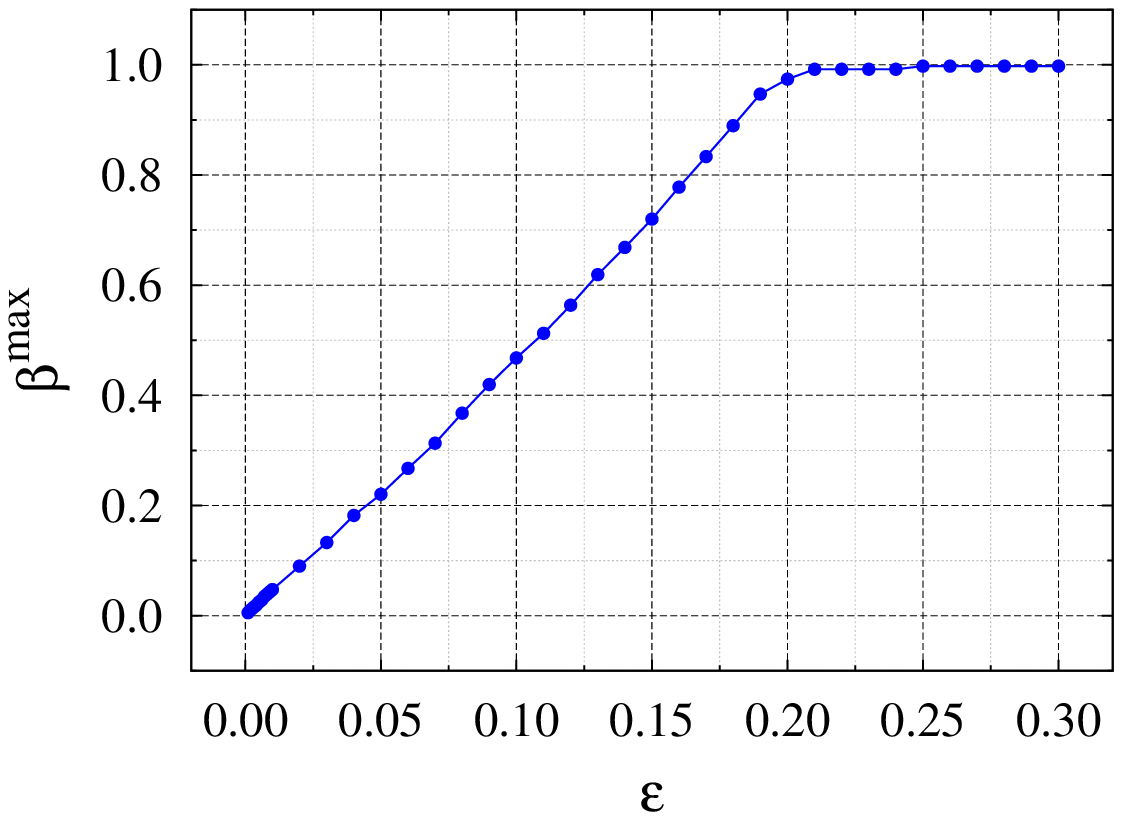}
\hspace*{1cm}
\includegraphics[width=0.40\textwidth]{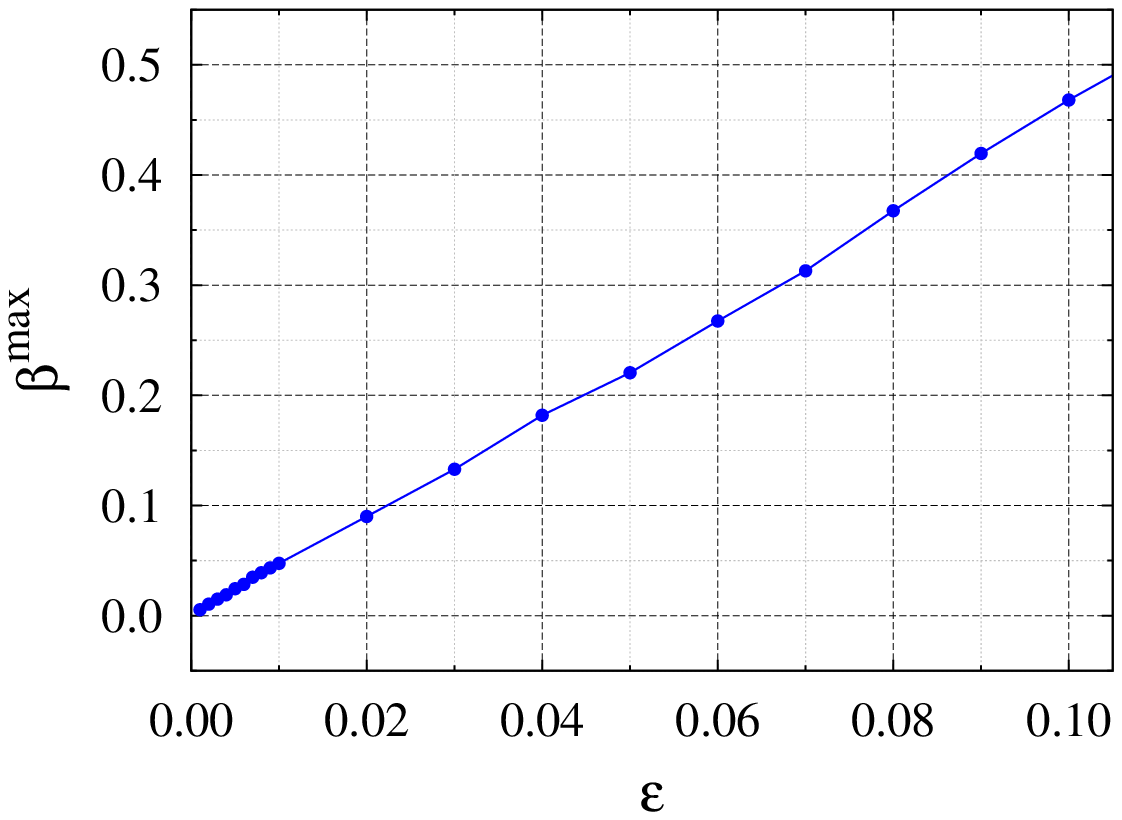}
\caption{The dependence of the maximal value of parameter $\beta$ on
precision $\varepsilon$ ranging from 0 to 0$''$.3 (\emph{left}) and
from 0 to 0$''$.1 (\emph{right}).}
\label{fig05}
\end{figure*}

\begin{figure*}[ht!]
\centering
\includegraphics[width=0.40\textwidth]{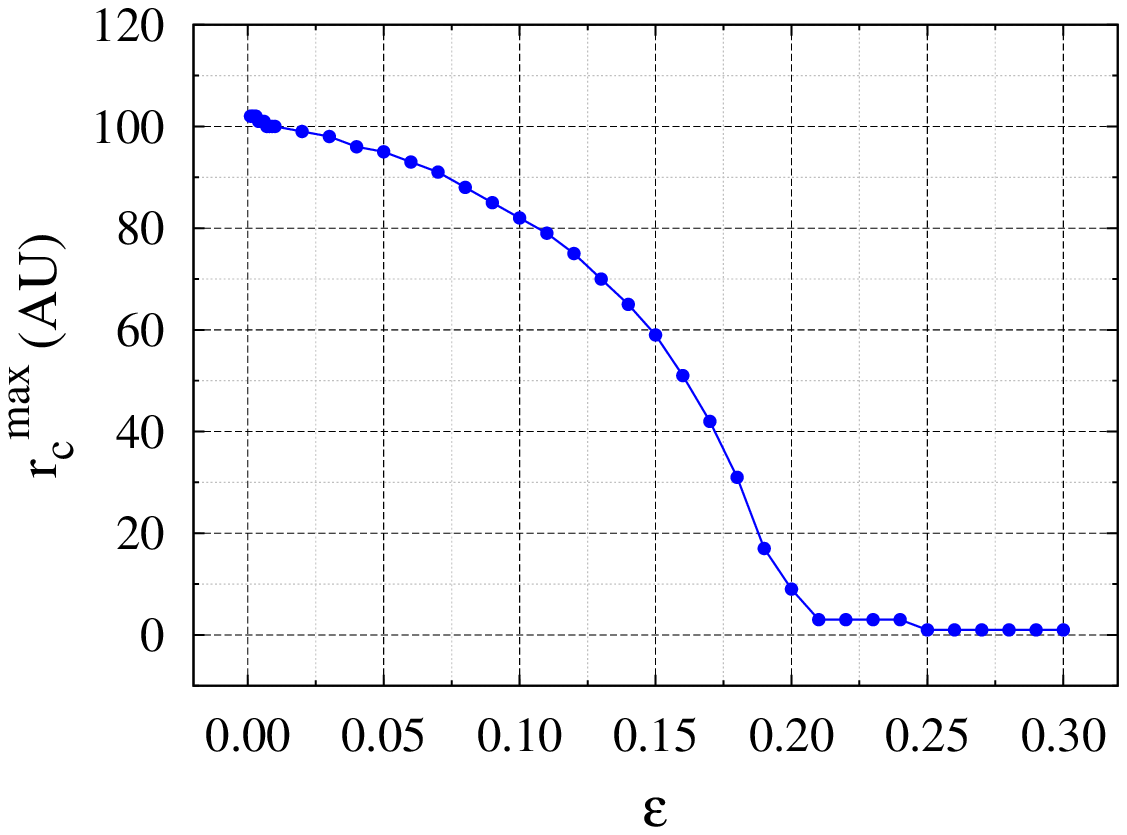}
\hspace*{1cm}
\includegraphics[width=0.40\textwidth]{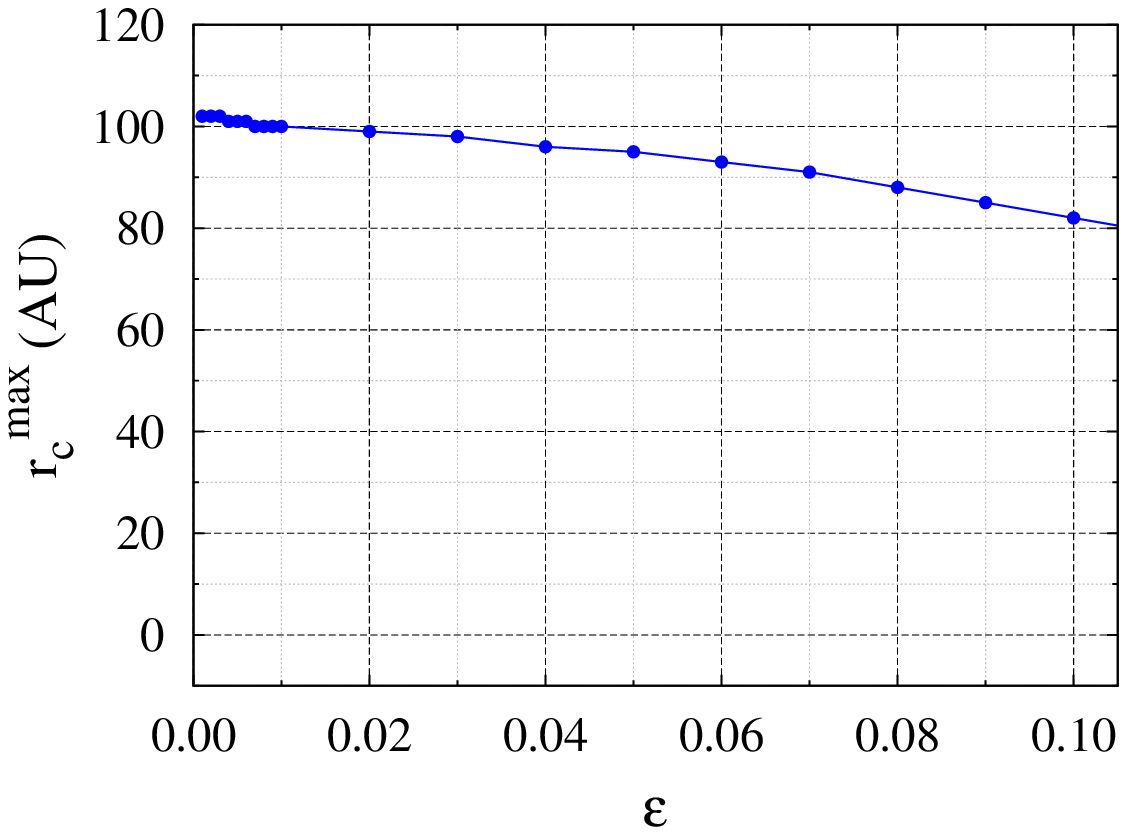}
\caption{The dependence of the $r_c^{max}$ on precision
$\varepsilon$ ranging from 0 to 0$''$.3 (\emph{left}) and from 0 to
0$''$.1 (\emph{right}).}
\label{fig06}
\end{figure*}

\begin{figure*}[ht!]
\centering
\includegraphics[width=0.40\textwidth]{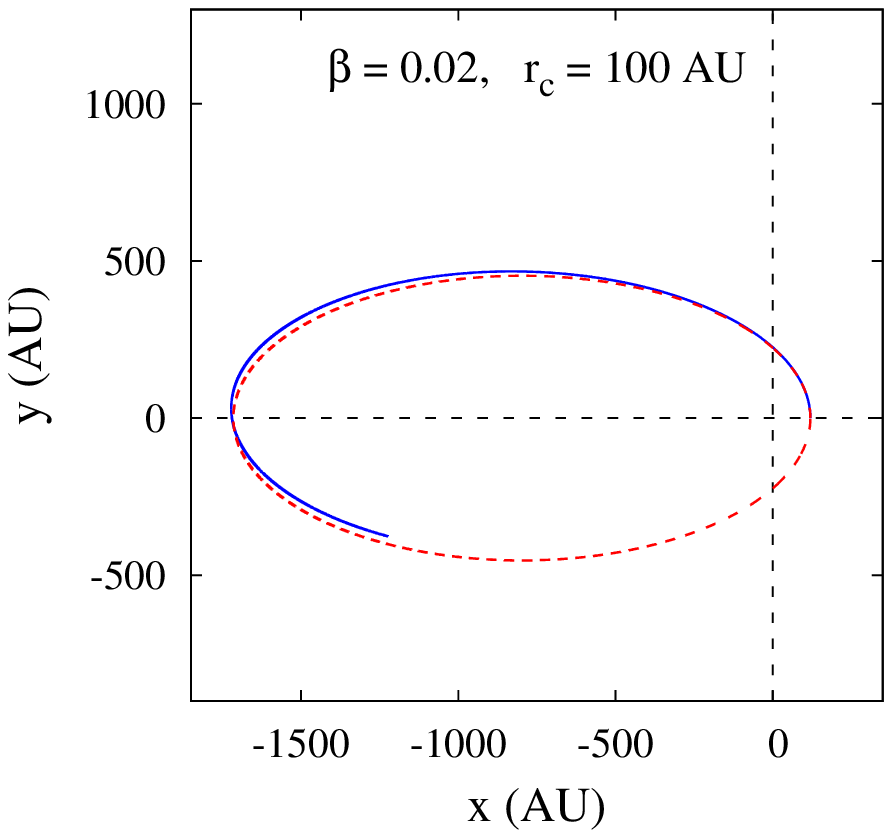}
\hspace*{1cm}
\includegraphics[width=0.40\textwidth]{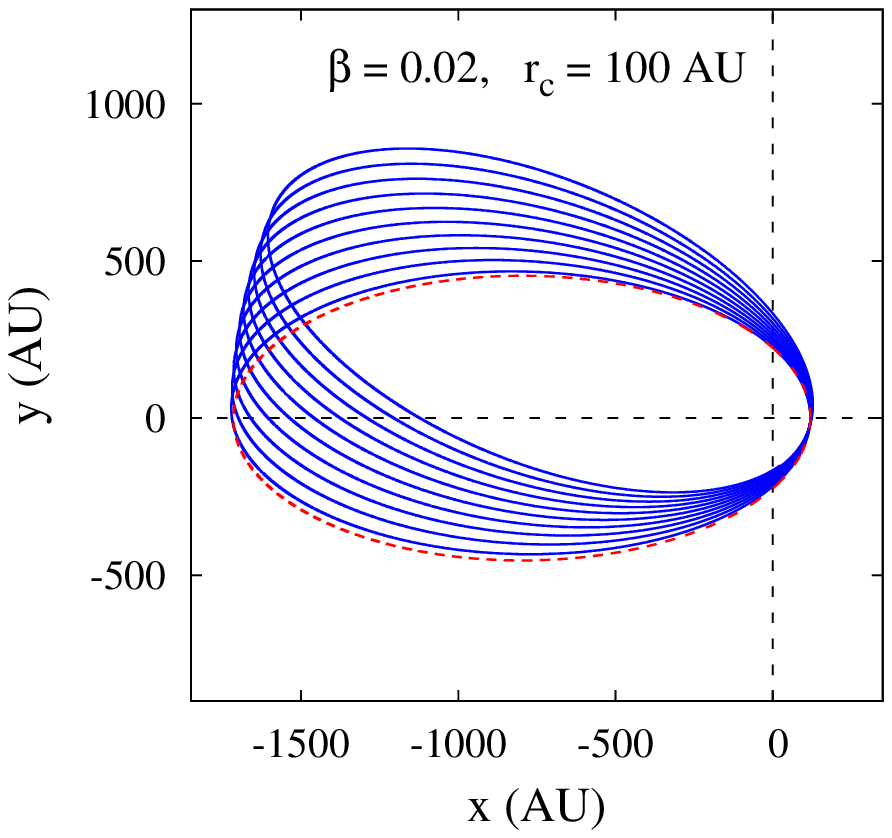}
\caption{The orbits of S2-like star around massive black hole in
$R^n$ gravity (blue solid line) and in Newtonian gravity (red dashed
line) for $r_c$ = 100 AU and $\beta$ = 0.02 during 0.8 periods
(\emph{left}) and 10 periods (\emph{right}).}
\label{fig07}
\end{figure*}

\begin{figure*}[ht!]
\centering
\includegraphics[width=0.85\textwidth]{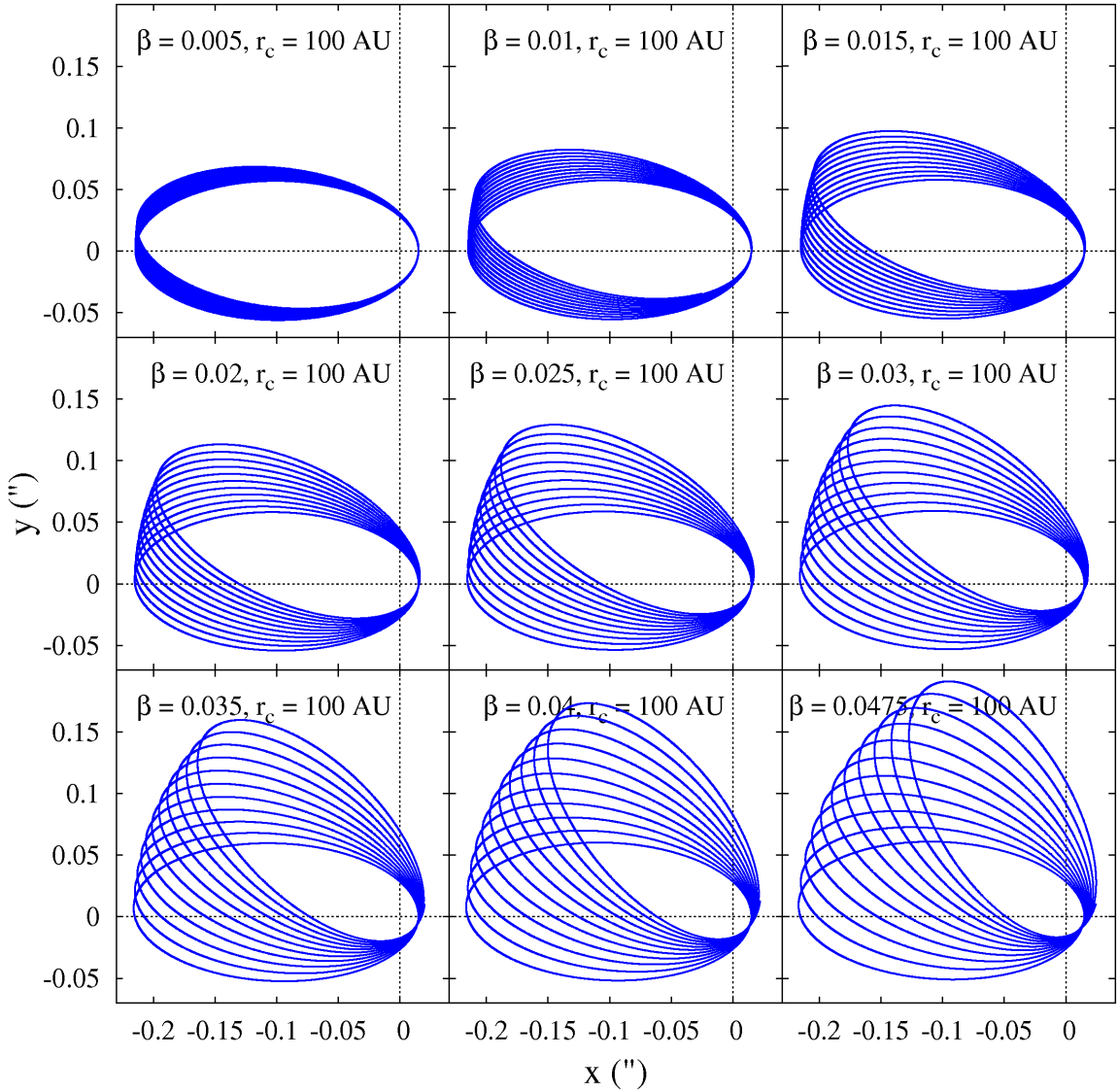}
\caption{The orbital precession of an S2-like star around massive
black hole located at coordinate origin in $R^n$ gravity for $r_c$ =
100 AU and these nine values of parameter $\beta$: 0.005, 0.01,
0.015, 0.02, 0.025, 0.03, 0.035, 0.04, 0.0475.}
\label{fig08}
\end{figure*}

\begin{figure}[ht!]
\centering
\includegraphics[width=0.45\textwidth]{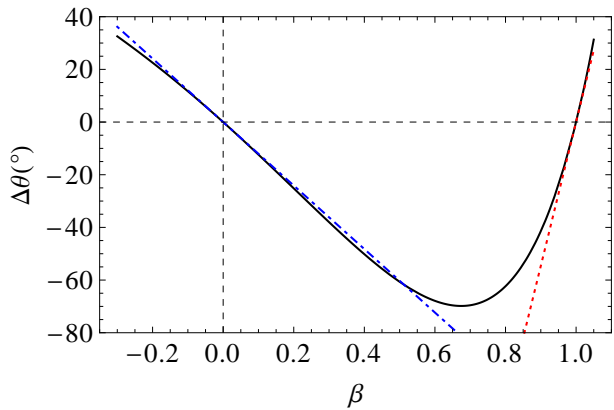}
\caption{The exact expression for precession angle $\Delta \theta$ in $R^n$ gravity (in
degrees) as a function of the parameter $\beta$ (black solid line) and two its
approximations: for $\beta \approx$ 0 (blue dash-dotted line) and for $\beta \approx$ 1
(red dotted line). The other parameters correspond to the case of S2-like star: $a = $ 919 AU,
$e = 0.87$ and $r_c$ = 100 AU.}
\label{fig09}
\end{figure}

\begin{figure*}[ht!]
\centering
\includegraphics[width=0.45\textwidth]{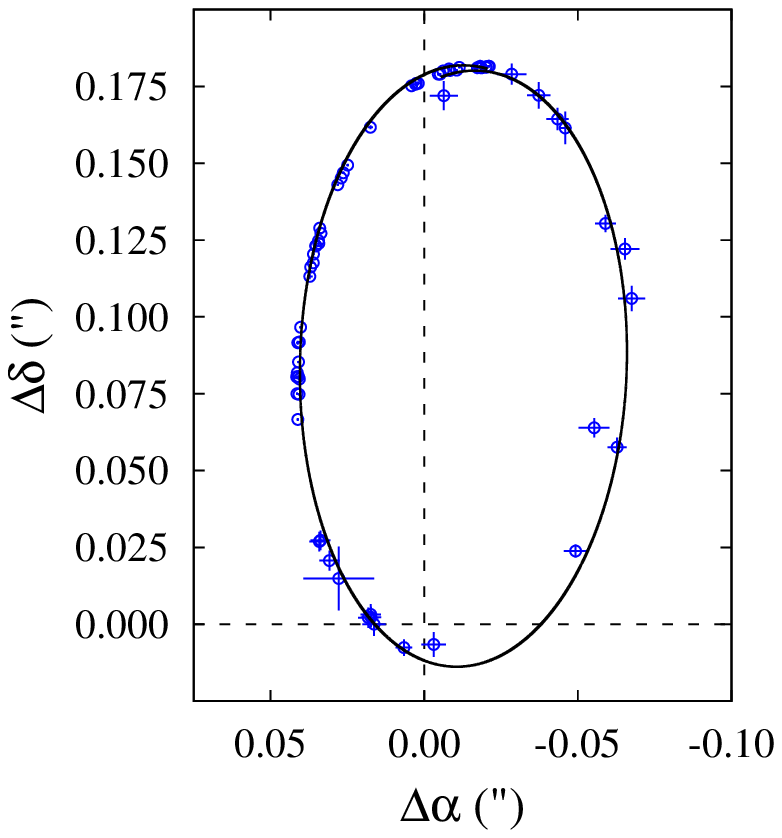}
\hspace*{0.8cm}
\includegraphics[width=0.45\textwidth]{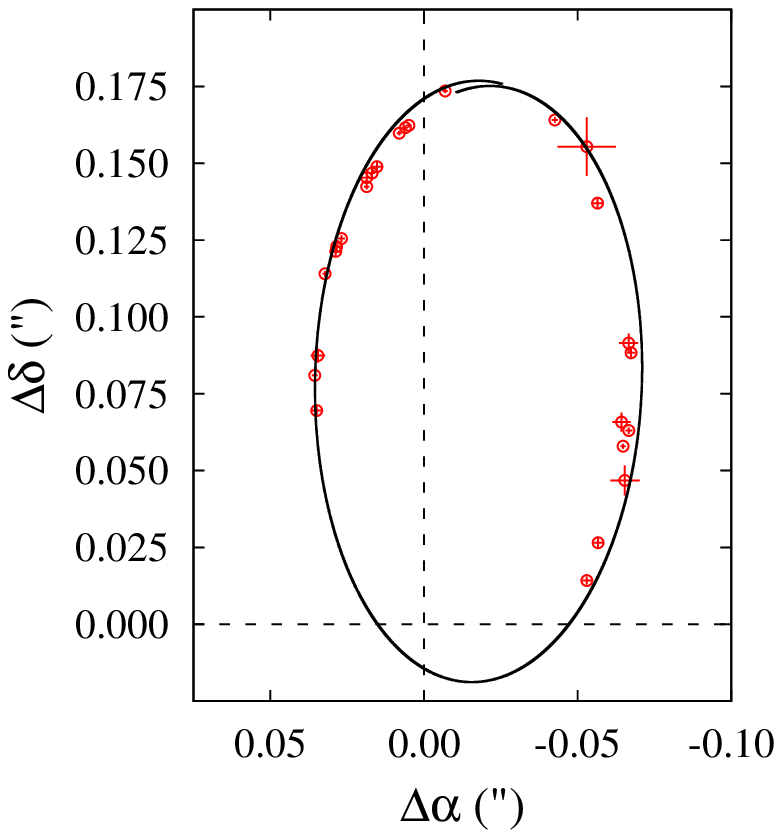}
\caption{The fitted orbit of S2 star around massive black hole in
$R^n$ gravity for $r_c$ = 100 AU and $\beta$ = 0.01 (black solid
lines in both panels). The NTT/VLT astrometric observations are
presented in the left panel by blue circles, while the Keck
measurements are denoted by red circles in the right panel.}
\label{fig10}
\end{figure*}

\begin{figure*}[ht!]
\centering
\includegraphics[width=0.49\textwidth]{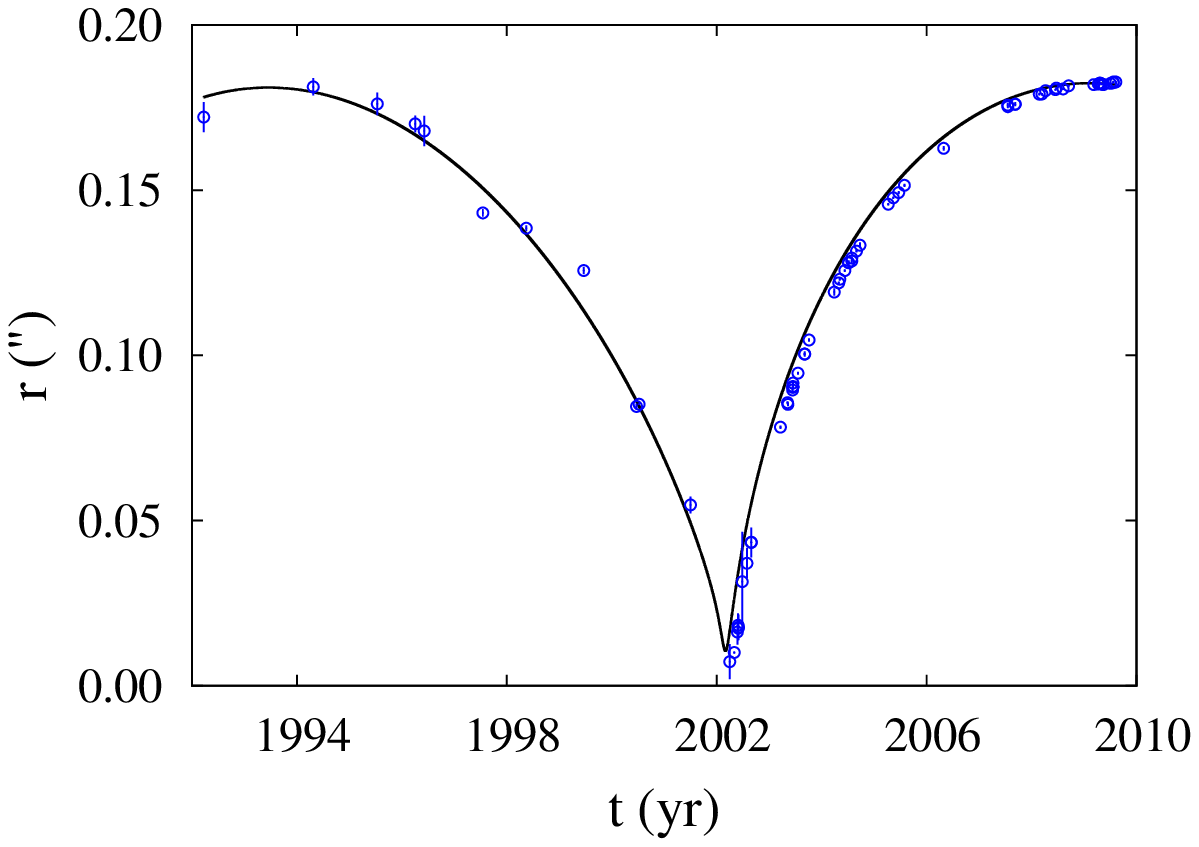}
\includegraphics[width=0.49\textwidth]{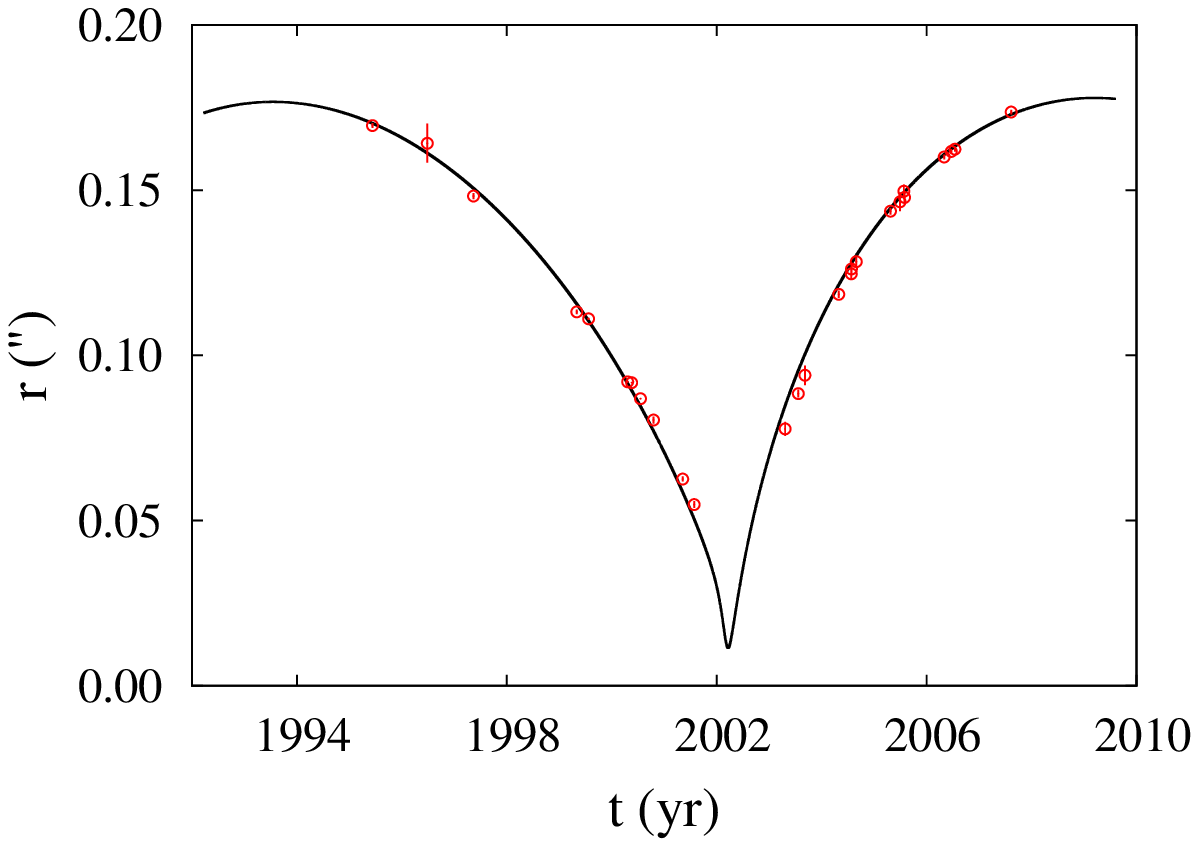}
\caption{Comparison between the fitted (black solid lines) and
measured (open circles) distances of the S2 star from black hole in
the case of NNT/VLT (left) and Keck (right) observations.}
\label{fig11}
\end{figure*}

\begin{figure}[ht!]
\centering
\includegraphics[width=0.49\textwidth]{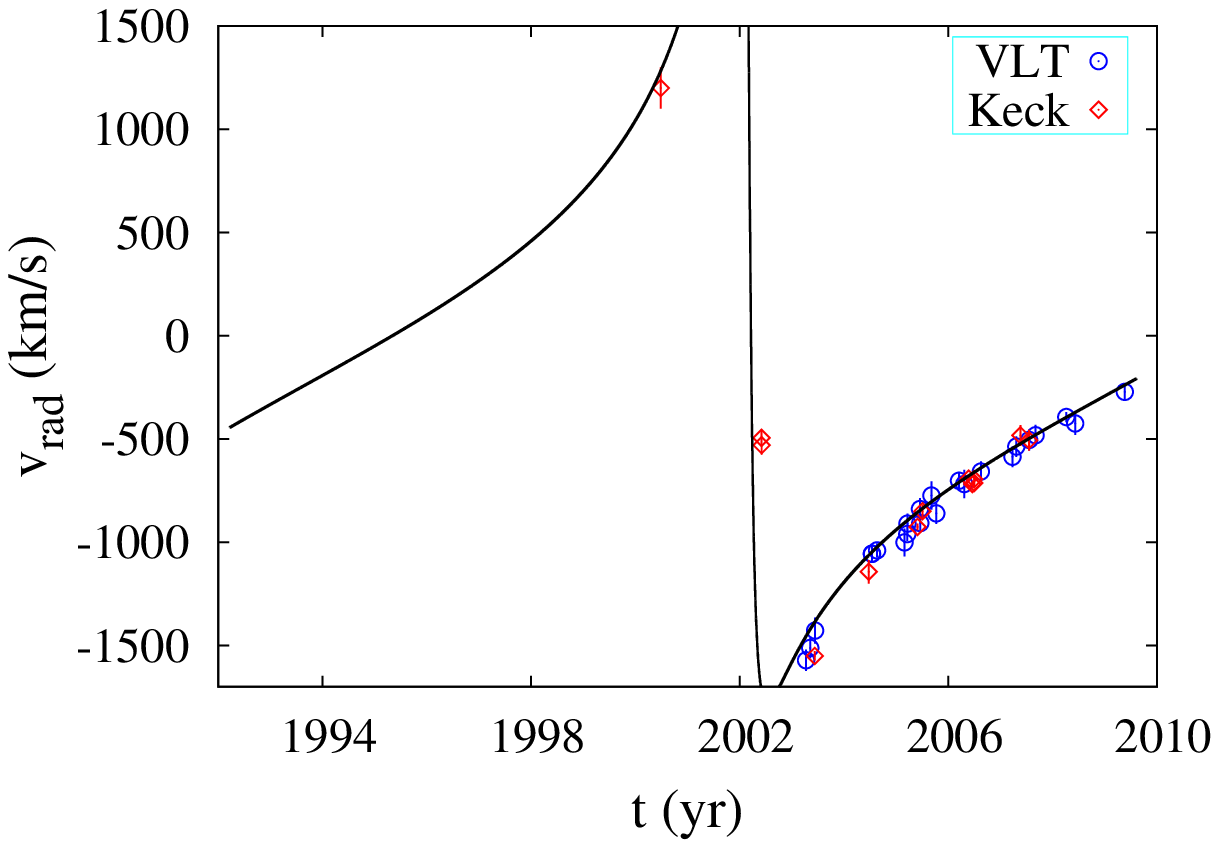}
\caption{Comparison between the fitted (black solid line)
and measured radial velocities for the S2 star. Measured velocities
are labeled with open circles (VLT data) and open rhombuses (Keck
data).}
\label{fig12}
\end{figure}

\section{Theory}

$R^n$ gravity belongs to power-law fourth-order theories of gravity
obtained by replacing the scalar curvature $R$ with $f(R)$ = $f_0
R^n$ in the gravity Lagrangian \cite{capo06,capo07}. As a result, in
the weak field limit \cite{clif05}, the gravitational potential is
found to be \cite{capo06,capo07}:

\begin{equation}
\Phi \left( r \right) = -\dfrac{GM}{2r}\left[ {1 + \left( {\dfrac{r}{r_c}} \right)^\beta} \right],
\label{equ01}
\end{equation}

\noindent where $r_c$ is an arbitrary parameter, depending on the
typical scale of the considered system and $\beta$ is a universal
parameter:

\begin{equation}
\beta = \dfrac{12 n^2 - 7n - 1 - \sqrt{36 n^4 + 12 n^3 - 83 n^2 + 50n + 1} }{6 n^2 - 4n + 2}
\label{equ02}.
\end{equation}

This formula corresponds to a modification of the gravity action in
the form:

\begin{equation}
A = \int {d^4 x \sqrt{-g} \left( f\left( R \right) + L_m \right)},
\label{equ03}
\end{equation}

\noindent where $f(R)$ is a generic function of the Ricci scalar
curvature and $L_m$ is the standard matter Lagrangian.

For $n$ = 1 and $\beta$ = 0 the $R^n$ potential reduces to the
Newtonian one, as expected. Parameter $\beta$ controls the shape of
the correction term and is related to $n$ which is part of the
gravity Lagrangian. Since it is the same for all gravitating
systems, as a consequence, $\beta$ must be the same for all of them
and therefore it is universal parameter \cite{capo07}. The parameter
$r_c$ is the scalelength parameter and is related to the boundary
conditions and the mass of the system \cite{capo07}.

\section{Results}

\subsection{Orbits of S2-like stars and parameters of $R^n$ gravity}

In order to study the effects of $R^n$ gravity on motion of the star
S2, we performed two-body calculations of its orbit in the $R^n$
potential (Eq.(\ref{equ01})) during two periods. We assumed the
following input parameters taken from the paper of Zakharov et al.
\cite{zakh07}: orbital eccentricity of S2-like star $e$ = 0.87,
major semi-axis $a$ = 919 AU, mass of S2-like star $M_\star = 1\
M_\odot$, mass of central black hole $M_{BH}$ = 3.4 $\times10^6
M_\odot$ (where $M_\odot$ is solar mass) and orbital period of
S2-like star is 15 years. We calculated S2-like star orbit during
two periods using Newtonian and $R^n$ potentials. We also
investigated the constraints on the parameters $\beta$ and $r_c$ for
which the deviations between the S2-like stars orbits in the $R^n$
gravity potential (Eq.(\ref{equ01})) and its Keplerian orbit will
stay within the maximum precision of the current instruments (about
10 mas), during one orbital period.

In Fig. \ref{fig01} we presented trajectory of S2-like star around
massive black hole in $R^n$ gravity (blue solid line) and in
Newtonian gravity (red dashed line) for $r_c$ = 100 AU and for the
following nine values of parameter $\beta$: 0.005, 0.01, 0.015,
0.02, 0.025, 0.03, 0.035, 0.04, 0.0475. The black hole is assumed to
be located at coordinate origin. We fixed a value of parameter $r_c$
on 100 AU because this value corresponds to maximal value of
parameter $\beta$ in the parameter space (see Fig. \ref{fig03}) and
varied values of parameter $\beta$. All 9 presented orbits fulfill
request that $R^n$ orbit and corresponding Newtonian orbit differ
less then 10 mas, (i.e. within the maximum precision of the current
observations) during one orbital period. We can see that if
parameter $\beta$ increases $R^n$ orbit differs more from the
corresponding Newtonian orbit since the precession angle becomes
larger. This indicates that the value of $\beta$ should be small, as
inferred from Solar system data \cite{zakh06}, and in contrast to
the value $\beta$ = 0.817 (obtained by \cite{capo07} which gives
excellent agreement between theoretical and observed rotation
curves). In the future, with improvements in observational
facilities the precision on constrains on values of parameters
$\beta$ and $r_c$ will increase, as well as the accuracy of the S2
orbit.

The corresponding distances between the S2-like star and black hole
as a function of time for the same values of parameters $r_c$ and
$\beta$ as in the Fig. \ref{fig01} are presented in Fig.
\ref{fig02}. There is an additional requirement on parameter space:
period of S2-like star orbit has to remain $\approx$15 $\pm$ 0.2 yr.
Like in previous case, with increasing observational accuracy of
period the precision on constraints on values of parameters $\beta$
and $r_c$ will also increase.

In Fig. \ref{fig03} we presented the parameter space for $R^n$
gravity under constrain that, during one orbital period, S2-like
star orbits under $R^n$ gravity differ less than $\varepsilon$ from
their orbits under Newtonian gravity for 10 values of parameter
$\varepsilon$: 0.001, 0.002, 0.003, 0.004, 0.005, 0.006, 0.007,
0.008, 0.009 and 0$''$.01. For $\varepsilon$ = 0$''$.01, we can see
that the maximal value of $\beta$ is 0.0475, and the corresponding
$r_c$ is 100 AU. That is why we investigate combinations $\beta$
$\leq$ 0.0475 and $r_c$ = 100 AU. This study is important because
with improvements in observational facilities the precision of
$\varepsilon$ will increase.

In Fig. \ref{fig04} we presented the parameter space for $R^n$
gravity under constraint that, during one orbital period, S2-like
star orbits in $R^n$ gravity differ less than $\varepsilon$ from the
corresponding orbits in Newtonian gravity for 12 values of parameter
$\varepsilon$: 0.01, 0.02, 0.03, 0.04, 0.05, 0.06, 0.07, 0.08, 0.09,
0.1, 0.11 and 0$''$.12. Although these values are unrealistic, since
they are larger then the current observational resolutions, they can
be used to obtain tendency of $\beta$ and $r_c$ dependences on
$\varepsilon$.

We want to find how the constraints on values of parameters $\beta$
and $r_c$ will change with increasing of $\varepsilon$. From Figs.
\ref{fig03} and \ref{fig04} we can see that smaller values of
$\varepsilon$ will significantly reduce parameter space of $R^n$
gravity within the given precision.

In Fig. \ref{fig05}(left and right) we presented dependence of the
maximal value of parameter $\beta$ versus $\varepsilon$ for $0 \leq
\varepsilon \leq 0''.3$ and $0 \leq \varepsilon \leq 0''.1$,
respectively. We can see that in the region of interest the
dependence of the maximal value of parameter $\beta$ versus values
of $\varepsilon$ is almost strictly linear.

In Fig. \ref{fig06}(left and right) we presented dependence of the
value of parameter $r_c^{max}$, that correspond to maximal value of
parameter $\beta$ (e.g. $\beta^{max}$), versus $\varepsilon$ for $0
\leq \varepsilon \leq 0''.3$ and $0 \leq \varepsilon \leq 0''.1$,
respectively. With decrease of $\varepsilon$ the value of parameter
$r_c^{max}$ increase and in the region of interest the value of
$r_c^{max}$ is near 100 AU.

The trajectories of S2-like star around massive black hole in $R^n$
gravity (blue solid line) and in Newtonian gravity (red dashed line)
are presented in Fig. \ref{fig07} (left and right) for $r_c$ = 100
AU and $\beta$ = 0.02 during 0.8 and 10 periods, respectively. We
can see that the precession of S2-like star orbit is in the
clockwise direction in the case when the revolution of S2-like star
is in counter clockwise direction.

In Figure \ref{fig08} we show calculated S2-like star orbits for 11
periods, assuming $r_c$ = 100 AU and these nine values of parameter
$\beta$: 0.005, 0.01, 0.015, 0.02, 0.025, 0.03, 0.035, 0.04, 0.0475.
We adopted value $r_c$ = 100 AU because it corresponds to the
largest allowed range of parameter $\beta$. We obtained that the
$R^n$ gravity causes periastron shifts which result in rosette
shaped orbits.

\subsection{Angle of orbital precession in $R^n$ gravity}

Talmadge et al. \cite{talm88} used post-Newtonian formalism of
metric theories of gravity in order to calculate perihelion
precession in a potential which deviates from Newtonian potential
only slightly. Adkins and McDonnell \cite{adki07} calculated the
precession of Keplerian orbits under the influence of arbitrary
central force perturbations. For some examples including the Yukawa
potential they presented the results using hypergeometric functions.
Schmidt \cite{schm08} calculated the perihelion precession of nearly
circular orbits in a central potential which has a form of modified
Newtonian potential. Since the S2-like star orbit is very eccentric,
and hence non-circular, we used approach from \cite{adki07} to
obtain analytical expression for precession angle in $R^n$ gravity.

Assuming a potential which does not differ significantly from
Newtonian potential, in our case $R^n$ gravitational potential, we
will derive formula for precession angle of the modified orbit,
during one orbital period. First step is to derive perturbing
potential from:

\begin{equation}
V(r) = \Phi \left( r \right) - {\Phi_N}\left( r \right)\begin{array}{*{20}{c}}
;&{{\Phi_N}\left( r \right) =  - \dfrac{{GM}}{r}}
\end{array}.
\label{equ04}
\end{equation}

\noindent Obtained perturbing potential is of the form:

\begin{equation}
V(r) = - \dfrac{GM}{2r}\left( {{{\left( {\dfrac{r}{r_c}} \right)}^\beta } - 1} \right),
\label{equ05}
\end{equation}

\noindent and it can be used for calculating the precession angle
according to the equation (30) from paper \cite{adki07}:

\begin{equation}
\Delta \theta = \dfrac{-2L}{GM e^2}\int\limits_{-1}^1 {\dfrac{z \cdot dz}{\sqrt{1 - z^2}}\dfrac{dV\left( z \right)}{dz}},
\label{equ06}
\end{equation}

\noindent where $r$ is related to $z$ via: $r = \dfrac{L}{1 + ez}$.
By differentiating the perturbing potential $V(z)$ and substituting
its derivative and expression for the semilatus rectum of the
orbital ellipse ($L = a\left( {1 - {e^2}} \right)$) in above
equation (\ref{equ06}), we obtain:

\begin{equation}
\begin{array}{lll}
\Delta \theta &=& \dfrac{\pi}{2}\beta \left( {\beta - 1} \right){\left(
{\dfrac{a\left( {1 - e^2} \right)}{r_c}} \right)^\beta }\times \\
&&\times{}_2{F_1}\left( {\dfrac{\beta + 1}{2},\dfrac{\beta + 2}{2};2;{e^2}}
\right),
\end{array}
\label{equ07}
\end{equation}

\noindent where $_2F_1$ is hypergeometric function. The graphical
presentation of the precession angle $\Delta \theta$ for S2-like
star orbit as a function of $\beta$ is given in Fig. \ref{fig09}
(black solid line). From this figure it can be seen that $\Delta
\theta$ is negative for all values of $\beta$ between 0 and 1, which
are of interest in the case of S2-like star orbit.

Exact expression (\ref{equ07}) is inappropriate for practical
applications. However, it can be easily approximated for
$\beta\approx 0$ and $\beta\approx 1$. In case of $\beta\approx 0$
expansion of Eq. (\ref{equ07}) in Taylor's series over $\beta$, up
to the first order, leads to the following expression for precession
angle:

\begin{equation}
\begin{array}{lll}
\Delta \theta &=& \dfrac{{\pi^{rad}}\beta \left( {\sqrt{1-e^2} - 1} \right)}{e^2} \\
&& \\
&=& \dfrac{{180^\circ}\beta \left( {\sqrt {1-e^2} - 1} \right)}{e^2}.
\end{array}
\label{equ08}
\end{equation}

Above expression in the case of S2-like star orbit is presented in
Fig. \ref{fig09} as a blue dash-dotted line. Similarly, expansion of
Eq. (\ref{equ07}) in power series for $\beta \approx$ 1, leads to
the following expression for precession angle (red dotted line in
Fig. \ref{fig09}):

\begin{equation}
\begin{array}{lll}
\Delta \theta &=& \dfrac{{{\pi^{rad}}a\left( {\beta - 1} \right)\left( {\sqrt {1 - e^2} - 1 + e^2} \right)}}{r_c e^2} \\
&& \\
&=& \dfrac{{{180^\circ }a\left( {\beta - 1} \right)\left( {\sqrt {1 - e^2} - 1 + e^2} \right)}}{r_c e^2}.
\end{array}
\label{equ09}
\end{equation}

One can expect that in general precession angle depends on semimajor
axis and eccentricity of the orbit (see e.g. Iorio \& Ruggiero 2008
\cite{iori08}), as well as on both potential parameters $\beta$ and
$r_c$. It is indeed case for $\beta \approx$ 1 in Eq. (\ref{equ09}).
But as it can be seen from formula (\ref{equ08}), the precession
angle in the case when $\beta$ is small ($\beta\approx 0$) depends
only on eccentricity and universal constant $\beta$ itself.

In order to test if the approximation from Eq. (\ref{equ08}) is
satisfactory in case of S2-like star, we derived its precession
angle in two ways:
\begin{itemize}
\item analytically from the approximative formula (\ref{equ08})
\item numerically from calculated orbits presented in Fig.
    \ref{fig08}.
\end{itemize}
Comparison of the obtained precession angles by these two methods is
presented in Table \ref{tab01}. As it can be seen from this table,
the approximative formula (\ref{equ08}) can be used for estimating
the precession angle for all values of $\beta$ from Fig.
\ref{fig08}.

Above analysis indicates that $R^n$ gravity results with retrograde
shift of S2-like star orbit. Rubilar and Eckart \cite{rubi01} showed
that the orbital precession can be due to relativistic effects,
resulting in a prograde shift, or due to a extended mass
distribution, producing a retrograde shift. We can conclude that
perturbing potential $V(r)$ has a similar effect as extended mass
distribution, since it produces a retrograde orbital shift.

\begin{table}[ht!]
\centering \caption{The numerically calculated by computer
simulation ($\Delta \theta^o$) and analytically calculated from Eq.
(\ref{equ08}) ($\Delta \theta^c$) values of precession angle of
S2-like star orbit (in degrees) as a function of universal constant
$\beta$ of $R^n$ gravity with parameter $r_c$ = 100 AU.}
\begin{ruledtabular}
\begin{tabular}{lcc}
\noalign{\smallskip}
$\beta$ & $\Delta \theta^o$ & $\Delta \theta^c$ \\
\noalign{\smallskip}
\hline
\noalign{\smallskip}
0.005 & -0.602 & -0.604 \\
\noalign{\smallskip}
\hline
\noalign{\smallskip}
0.01 & -1.203 & -1.209 \\
\noalign{\smallskip}
\hline
\noalign{\smallskip}
0.015 & -1.802 & -1.816 \\
\noalign{\smallskip}
\hline
\noalign{\smallskip}
0.02 & -2.400 & -2.425 \\
\noalign{\smallskip}
\hline
\noalign{\smallskip}
0.025 & -2.997 & -3.035 \\
\noalign{\smallskip}
\hline
\noalign{\smallskip}
0.03 & -3.592 & -3.647 \\
\noalign{\smallskip}
\hline
\noalign{\smallskip}
0.035 & -4.186 & -4.261 \\
\noalign{\smallskip}
\hline
\noalign{\smallskip}
0.040 & -4.779 & -4.876 \\
\noalign{\smallskip}
\hline
\noalign{\smallskip}
0.045 & -5.666 & -5.493 \\
\noalign{\smallskip}
\end{tabular}
\end{ruledtabular}
\label{tab01}
\end{table}

Since the precession has negative direction, as in the case of
extended mass distribution, the obtained results are useful for
testing if the precession due to extended dark matter enclosed into
orbit of S2-like star could be also explained by $R^n$ gravity. If
this is possible, it will exclude the need for dark matter
hypothesis. Therefore, if future and more precise observations of
bright stars near the Galactic Center will show a precession in the
negative direction, we have to conclude that  the phenomenon could
be caused by bulk distributions of stellar cluster or/and dark
matter in classical Newtonian (GR) gravity  or by $R^n$ gravity. On
the other hand, if there is no deviations from Newtonian (GR)
trajectories with an accuracy of observations, one could put
constraints on stellar cluster and dark matter distributions and on
parameters of $R^n$ gravity if we adopt the theory to fit
observational data.

\subsection{Comparison between the theoretical results and
observations}

Here we compare the obtained theoretical results for S2-like star
orbits in the $R^n$ potential with two independent sets of
observations of the S2 star, obtained by New Technology
Telescope/Very Large Telescope (NTT/VLT), as well as by Keck
telescope \citep[see Fig. 1 in][]{gill09a}, which are publicly
available as the supplementary on-line data to the electronic
version of the paper \citep{gill09a}. However, all the above
two-body simulations in $R^n$ gravity potential resulted with the
true orbits of S2-like stars, i.e. the simulated positions of
S2-like stars presented in Figs. \ref{fig01}, \ref{fig02},
\ref{fig07} and \ref{fig08} are in their orbital planes. Therefore,
in order to compare them with observed positions, the first step is
to project them to the observer's sky plane, i.e. to calculate the
corresponding apparent orbits. From the theory of binary stars it is
well known that any point $(x, y)$ on the true orbit could be
projected into the point $(x^c, y^c)$ on the apparent orbit
according to \cite[see e.g.][]{aitk18,smar30}:
\begin{equation}
x^c=l_1 x+l_2 y ,\hspace*{0.5cm} y^c=m_1 x+m_2 y ,
\label{equ10}
\end{equation}
\ where the expressions for $l_1, l_2, m_1$ and $m_2$ depend on
three orbital elements ($\Omega$ - longitude of the ascending node,
$\omega$ - longitude of pericenter and $i$ - inclination)
\citep{aitk18,smar30}. One should take into account that in the case
of orbital precession $\omega$ is a function of time, and therefore
should be in general treated as an adjustable parameter during the
fitting procedure. However, the previously mentioned theoretical and
observational results showed that in the case of S2-like stars this
precession is most likely very small, and hence we assumed $\omega$
as a constant when projecting true positions to their corresponding
apparent values. For that purpose we used the following Keplerian
orbital elements from \cite{gill09a}: $i=134^{\circ}.87$,
$\Omega=226^{\circ}.53$ and $\omega=64^{\circ}.98$. Besides, our
previous theoretical results indicated that the most likely value of
the scale parameter is $r_c\approx 100$ AU, and  we adopted that
value in order to reduce the number of free parameters when fitting
the observations.

We fitted the observed orbits of S2 star using the following
procedure:
\begin{enumerate}
\item initial values for S2 star true position $(x_0, y_0)$,
    orbital velocity $(\dot{x}_0, \dot{y}_0)$ and the
    parameter $\beta$ of $R^n$ gravity potential are specified;
\item the positions $(x_i, y_i)$ of the S2 star along its true orbit are
    calculated at the observed epochs using two-body simulations
    in the $R^n$ gravity potential, assuming that distance to
    the S2 star is $d_\star$ = 8.3 kpc and mass of central black
    hole $M_{BH}$ = 4.4 $\times10^6 M_\odot$ \cite{gill09a};
\item the corresponding positions $(x_i^c, y_i^c)$ along the
    apparent orbit are calculated using the expression
    (\ref{equ10});
\item the root mean square $(O-C)$ goodness of fit is estimated
    according the following expression:\newline $(O-C)_{rms} =
    \sqrt {\dfrac{{\sum\limits_{i = 1}^N {\left[ {{{\left(
    {x_i^o - x_i^c} \right)}^2} + {{\left( {y_i^o - y_i^c}
    \right)}^2}} \right]} }}{{2N}}},$ where $(x_i^o, y_i^o)$ is
    the $i$-th observed position, $(x_i^c, y_i^c)$ is the
    corresponding calculated position, and $N$ is the number of
    observations;
\item the values of the input parameters are varied and the
    procedure is repeated until the minimum of $(O-C)_{rms}$ is
    reached.
\end{enumerate}

The best fit is obtained for the following small value of the
universal constant: $\beta = 0.01$, in which case the corresponding
precession is around $-1^\circ$ (see Table \ref{tab01}). In Fig.
\ref{fig10} we present two comparisons between the obtained best fit
orbit for $\beta = 0.01$ in the $R^n$ gravity potential and the
positions of S2 star observed by NTT/VLT (left) and Keck (right).
The corresponding calculated distances of S2 star from massive black
hole are shown in Fig. \ref{fig11}. Astrometric data for the S2 star
orbit are presented by blue dots (NTT/VLT measurements) and by red
dots (Keck measurements). As one can see from these figures, there
is a good agreement between the theoretical orbit and NTT/VLT
observations. In case of Keck measurements, we had to move the
origin of the coordinate system with respect to the both axes for 5
mas, in order to get reasonable fit. We made this correction
following the suggestion from \cite{gill09a}, where it was necessary
in order to combine the two data sets. After that we also achieved
the satisfying agreement between the same fitted orbit and the both
NTT/VLT and Keck data sets, in spite the fact that both groups
obtained slightly different orbital elements, distance to the S2
star, as well as mass of central black hole
\cite{ghez03,ghez08,gill09a}.

In order to obtain the orbital elements of S2 star both, NTT/VLT and
Keck groups, fitted their observations with Keplerian orbits, but at
the same time, they had to allow that the position of the center of
mass varies as a function of time, i.e. they had to introduce the
orbital precession. As it can be seen from Fig. \ref{fig10}, the
orbit of S2 is not closed in vicinity of its apocenter, which
clearly shows that the orbital precession is a natural consequence
of $R^n$ gravity. Moreover, by comparing the arcs of orbit near the
apocenter with the corresponding results presented in Fig. 1 from
\cite{gill09a}, one can see that their curvatures are different,
which indicates the opposite directions of precession in these two
cases. Therefore, the future more precise observations of S2 star
positions near its apocenter could have a decisive role in verifying
or disproving the validity of $R^n$ gravity near the Galactic
Center.

We also made a comparison between the fitted and measured radial
velocities for the S2 star (see Fig. \ref{fig12}). The well known
expression for radial velocity in polar coordinates $r$ and $\theta$
is  \cite[see e.g.][]{aitk18}:
\begin{equation}
v_{rad} = \sin i \left[ \sin(\theta + \omega) \cdot \dot{r} + r
\cos (\theta + \omega) \cdot \dot{\theta} \right] .
\label{equ12}
\end{equation}
However, we used the corresponding expression in rectangular
coordinates $x=r\cos\theta$ and $y=r\sin\theta$ to calculate the
fitted radial velocities:
\begin{equation}
\begin{array}{ll}
v_{rad} =  & \dfrac{\sin i}{\sqrt{x^2 + y^2}} \left[ \sin(\theta + \omega)
\cdot(x \dot{x} + y \dot{y}) + \right. \\
& \\
 & \left. + \cos (\theta + \omega) \cdot (x \dot{y} - y \dot{x}) \right], \\
\end{array}
\label{equ13}
\end{equation}
\noindent where $\theta = \arctan \dfrac{y}{x}$. As it can be seen
from Fig. \ref{fig12}, the agreement between our theoretical
predictions and the observations is also satisfactory.

Although the both observational sets indicate that the orbit of S2
star most likely is not a Keplerian one, the nowadays astrometric
limit of around 10 mas is not sufficient to unambiguously confirm
such claim. We hope that in the future, it will be possible to
measure the stellar positions with astrometric errors several times
smaller than errors of current observations.

\section{Conclusions}

In this paper S2-like star orbit has been investigated in the
framework of fourth order gravity theory. Using the observed
positions of S2 star we put new constraints on the parameters of
this class of gravity theories. We confirmed that these parameters
must be very close to those corresponding to the Newtonian limit of
the theory. For parameter $\beta$ approaching to zero, we recover
the value of the Keplerian orbit for S2 star. Also, we performed
two-body calculations of its orbit in the $R^n$ potential. The
obtained results showed that, in contrast to General Relativity,
$R^n$ gravity gives retrograde direction of the precession of the S2
orbit, like in the case when it is caused by an extended matter
concentration in Newtonian potential.

Despite the excellent agreement between theoretical and observed
rotation curves obtained by Capozziello and coworkers \cite{capo07}
for $R^n$ parameter $\beta$ (the slope $n$ of the gravity Lagrangian
is set to the value $n$ = 3.5 giving $\beta$ = 0.817), our findings
indicate that for $\varepsilon$ = 0$''$.01 maximal value of $\beta$
is 0.0475, i.e. $\beta$ is less or equal than 0.0475, and our
fitting indicated that optimal value for $\beta$ is around 0.01.
Therefore, $R^n$ gravity in this form may not represent a good
candidate to solve both the dark energy problem on cosmological
scales and the dark matter one on galactic scales using the same
value of parameter $\beta$. But this theory has its own benefits in
explaining orbits of the stars and solar system data.

For today astrometric limit of around 10 mas for S2 star orbit,
within that limit one can not say for sure that S2 star orbit really
deviates from the Newtonian case, i.e. we have to stress that at the
moment observations are in agreement with the Newtonian point-like
potential for the Galactic Center. Therefore the observations and
their theoretical analysis give us one of the best cases to discuss
departures from the standard GR plus stellar cluster and dark matter
(as it was done in our papers and papers of other authors) or to
analyze an opportunity to get constraints on alternative theories
observing trajectories of S2-like stars. The newest astrometric data
for the star S2 of NTT/VLT measurements and Keck measurements show
the Keplerian orbit fits for the respective data set, do not yield
closed ellipses. Maybe this represents really small deviation of S2
star orbit from the Newtonian case and for more sure conclusion we
need astrometric errors several times smaller than these errors, but
we compared these data with S2 star orbit obtained using $R^n$
gravity potential.

We can conclude that additional term in $R^n$ gravity compared to
Newtonian gravity has a similar effect like extended mass
distribution and produce a retrograde shift, that results in rosette
shaped orbits.


\begin{acknowledgments}
This research is part of the project 176003 ''Gravitation and the
Large Scale Structure of the Universe'' supported by Ministry of
Education and Science of the Republic of Serbia.
\end{acknowledgments}


\end{document}